\documentclass[11pt,a4paper]{article}
\usepackage[hyperref]{emnlp2020}
\usepackage{times}
\usepackage{latexsym}
\usepackage{todonotes}
\usepackage{tabu}
\usepackage{booktabs}
\usepackage{multirow, tabularx}
\usepackage[capitalise]{cleveref}

\usepackage{subfig}
\usepackage{enumitem}
\usepackage{microtype}
\newcommand\blfootnote[1]{%
  \begingroup
  \renewcommand\thefootnote{}\footnote{#1}%
  \addtocounter{footnote}{-1}%
  \endgroup
}

\aclfinalcopy 

\title{Embedding-based Zero-shot Retrieval through Query Generation}

\author{Davis Liang $^{\spadesuit, \dagger}$\quad Peng Xu $^{\spadesuit, \dagger}$\quad Siamak Shakeri $^\diamondsuit$\thanks{\ \ Work done at Amazon.}\quad Cicero Nogueira dos Santos $^{\spadesuit}$\\ \texttt{\{liadavis, pengx, cicnog\}@amazon.com, siamaks@google.com} \vspace{-3mm} \AND Ramesh Nallapati$^{\spadesuit}$\quad Zhiheng Huang$^{\spadesuit}$\quad Bing Xiang$^{\spadesuit}$\\ \texttt{\{rnallapa, zhiheng, bxiang\}@amazon.com} \\ \\ AWS AI Labs$^\spadesuit$, Google$^\diamondsuit$\ }

\begin{document}
\maketitle
\begin{abstract}
Passage retrieval addresses the problem of locating relevant passages, usually from a large corpus, given a query. In practice, lexical term-matching algorithms like BM25 are popular choices for retrieval owing to their efficiency. However, term-based matching algorithms often miss relevant passages that have no lexical overlap with the query and cannot be finetuned to downstream datasets. In this work, we consider the embedding-based two-tower architecture as our neural retrieval model. Since labeled data can be scarce and because neural retrieval models require vast amounts of data to train, we propose a novel method for generating synthetic training data for retrieval. Our system produces remarkable results, significantly outperforming BM25 on 5 out of 6 datasets tested, by an average of 2.45 points for Recall@1. In some cases, our model trained on synthetic data can even outperform the same model trained on real data.

\blfootnote{$\dagger$ Corresponding Authors}
%
\end{abstract}
\section{Introduction}
\begin{table*}[t!]
\begin{minipage}{0.99\linewidth}
	\centering
	\footnotesize
    \begin{tabu}{p{0.57\textwidth}p{0.43\textwidth}}
    \toprule
    \textbf{Original Passage} & \textbf{Synthetic Queries} \\
    \midrule
Medical errors affect one in 10 patients worldwide . One extrapolation suggests that 180,000 people die each year partly as a result of iatrogenic injury . One in five Americans ( 22 \% ) report that they or a family member have experienced a medical error of some kind . The World Health Organization registered 14 million new cases and 8.2 million cancer-related deaths in 2012 . It estimated that the number of cases could increase by 70 \% through 2032 . As the number of cancer patients receiving treatment increases , hospitals around the world are seeking ways to improve patient safety , to emphasize traceability and raise efficiency in their cancer treatment processes .  
                  & how many cancer deaths are preventable \newline
                  how many people die from cancer every year \newline
                  number of deaths from cancer due to medical errors \newline
                  how many deaths are caused by miscommunication     \newline
                  how many americans a year experience a medical error \\
\midrule
For the rest of her life , she kept a cloth tied to her eyes and thus deprived herself of the power of sight . At certain critical junctures , she gave advice to her husband which was impeccable from a moral standpoint ; she never wavered in her adherence to dharma ( righteousness ) , even to a very bitter end . She was fated to witness the death of all her hundred sons within the space of 18 days , during the Great War between them and their cousins ; she also curses the lord Krishna when she was full of sorrow on the death of her 100 children that his vansh ( Clan ) would also be destroyed in the same manner as that of her . &
meaning of dharma \newline
is dharma righteous \newline
who said, dharma is righteousness \newline
why was vedanta so bitter \newline
why is there a garment tied to your eyes \\

    \bottomrule
    \end{tabu}
\end{minipage}
\caption{Examples of synthetic queries from \textsc{WikiGQ}.}
\label{tab:examples}
\end{table*} 
We consider the large-scale ad-hoc retrieval problem---retrieving relevant documents from a large corpus, given a query. Algorithms such as BM25 \cite{robertson2009probabilistic}, based on lexical term-matching, are among the most enduring and well-weathered models in classic information retrieval (IR). In fact, they enjoy widespread use in state-of-the-art ranking systems, where typically a deep neural ranking model re-ranks the BM25 retrieval results \cite{nogueira2019passage, macavaney2019cedr,yilmaz2019applying,nogueira2020document}. 

However, lexical matching algorithms are unable to capture semantic similarities not involving lexical overlap, are not trainable on target datasets, and cannot fully leverage recent advances in pretrained representations \cite{devlin2018bert, liu2019roberta}. These limitations make term-matching algorithms sub-optimal for passage ranking or question answering \cite{lee2019latent,guu2020realm,karpukhin2020dense}.

\begin{figure}[t!]
  \centering
  \includegraphics[width=1.0\columnwidth]{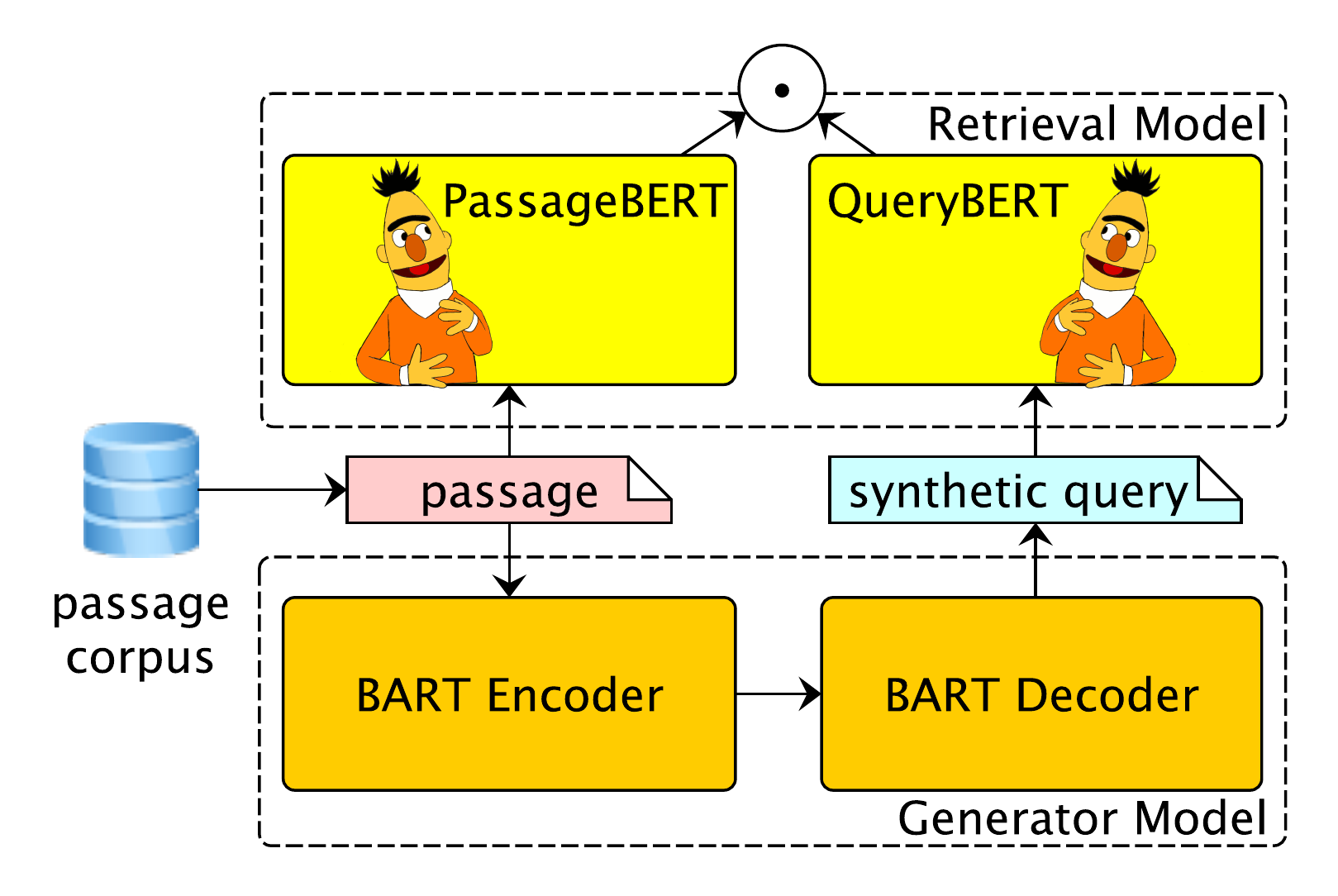}
  \caption{Synthetic queries, generated by a sequence-to-sequence model, are used to train our two-tower retrieval architecture.}
  \label{fig:bertonly}
\end{figure}

In this work, we consider the embedding-based two-tower architecture for neural retrieval. In this model, query and document embeddings are constructed from two encoders (i.e., ``towers") and the documents are ranked based on the similarity of these embeddings. Since document embeddings can be pre-computed, online retrieval is efficient with various approximate nearest-neighbor search algorithms \cite{aumuller2017ann}. The two-tower architecture, originated from the Siamese network \cite{bromley1994signature}, is particularly well-studied for the retrieval task \cite{severyn:2015,dossantos2015,das2016together,ma2019universal,cer2018universal,reimers2019sentence,lee2019latent,chang2020pre}.

As with all deep neural models, the two-tower architecture relies on the availability of a large amount of training data. However, for retrieval, obtaining labeled training data can often be expensive or even impossible. One possible solution exists with weak supervision; one can utilize noisy but cheap supervision signals such as query logs \cite{macavaney2017approach}, traditional IR results \cite{dehghani2017neural}, or an ensemble of multiple sources \cite{xu2019passage} to create training data. However, even these weakly supervised approaches rely on either an existing query set or external query log files, which are not always available. 

In this paper, we leverage synthetic queries generated from a large sequence-to-sequence (seq2seq) model for pretraining and unsupervised domain adaptation of the two-tower model. Specifically, we finetune BART \cite{lewis2019bart} on \textsc{MSMARCO} positive query-passage pairs to perform query generation (QG). Then, we construct a large-scale dataset by applying the BART QG model on English Wikipedia passages to generate synthetic query-passages pairs. We find that the BART-based QG model can generate surprisingly high-quality synthetic queries. For some datasets such as \textsc{Natural Questions}, training on these synthetic datasets can surpass the performance of training on the official training set. 

In the zero-shot setting, described in \cref{sec:results}, the Siamese model pretrained on synthetic Wikipedia queries significantly outperforms BM25 baselines on several datasets from both Wikipedia and non-Wikipedia domains. In addition, we apply our QG model on target domain datasets (e.g., \textsc{ANTIQUE}, \textsc{BioASQ}, \textsc{InsuranceQA}, etc.) to generate domain-specific synthetic data. We demonstrate that finetuning our retrieval models on these domain-specific synthetic queries can further improve performance.

Our contributions are as follows:
\begin{itemize}[topsep=2pt,itemsep=0pt,partopsep=0pt, parsep=2pt]
\item  We demonstrate a query generation method that can synthesize large amounts of high-quality data for the retrieval task. Finetuning on \textit{synthetic} queries generated from the target domain to perform \textit{unsupervised domain adaptation} can improve zero-shot performance on those target datasets. 
\item We achieve state-of-the-art performance on the ReQA benchmark.
\item We explore a variety of ablations, considering variations in embedding sizes, architecture choices, decoding methods, and more.
\end{itemize}

\begin{table*}[t!]
\begin{minipage}{1.0\linewidth}
	\centering
	\footnotesize
\begin{tabu}{@{}lccccc@{}}
\toprule
\textbf{Dataset}           & \textbf{Domain}   & \textbf{\# Passages} & \textbf{\# Evaluation queries} & \textbf{\# Train queries} & \begin{tabular}[c]{@{}c@{}}\textbf{Length of passages}\\ mean (std)\end{tabular} \\
\midrule
\textsc{WikiGQ}            & Wiki     & 22M        & -           & -   & 123.3 (32.5)                                                            \\
\textsc{Natural Questions} & Wiki     & 84,783      & 4,340       & 110,865   & 128.2 (74.2)                                                            \\
\textsc{TREC-CAR}          & Wiki     & 3.5M       & 195,659     & 584,461   & 92.9 (77.5)                                                             \\
\textsc{MSMARCO}           & non-Wiki & 8.8M       & 55,578      & 532,761   & 77.0 (30.9)                                                             \\
\textsc{InsuranceQA}       & non-Wiki & 27,288      & 1,625       & 10,391   & 121.3 (82.5)                                                            \\
\textsc{ANTIQUE}           & non-Wiki & 403,666     & 200         & 2,426   & 54.3 (74.8)                                                             \\
\textsc{BioASQ}            & non-Wiki & 20M        & 500         & 2,747   & 288.7 (130.6)                                                           \\
\midrule
\textsc{ReQA SQuAD}        & Wiki     & 97,707      & 87,599      & -   & 209.4 (79.7)
                                  \\
\textsc{ReQA NQ}           & Wiki     & 239,013     & 74,097      & -   & 193.2 (112.2)
                                  \\
\bottomrule
\end{tabu}
\end{minipage}
\caption{Summary of the datasets.}
\label{table:dataset}
\end{table*}

\section{Models}
In this section, we describe our methodology for query generation and its application to the passage retrieval task as illustrated in \cref{fig:bertonly}.

\subsection{Query Generator}
\label{ssec:bart}
We formulate query generation as a seq2seq task, where the input is a passage and the target output is a relevant query. 
Positive query-passage pairs, used to finetune our query generator, can be found in passage ranking datasets such as MSMARCO \cite{ nguyen2016ms} or extractive QA datasets such as \textsc{SQuAD} \cite{rajpurkar2016squad}. 
After finetuning the query generator, we apply it to perform query synthesis on arbitrary text corpora. We consider two use cases. First, we apply our generator to passages from English Wikipedia to create a large-scale synthetic pretraining dataset. Second, we apply the generator to the passages of target domain corpora 
and synthesize queries to create domain-specific synthetic data. Neither case requires human-annotated data in the target domain.

In this work, we adopt BART \cite{lewis2019bart}, a pretrained Transformer-based seq2seq model, as our query generator.

\subsection{Passage Retriever}
\label{ssec:bert}
BERT \cite{devlin2018bert} is a bidirectional transformer encoder model pretrained on English Wikipedia and BooksCorpus \cite{Zhu_2015_ICCV}. We leverage BERT-base as the foundation for our retrieval models. On top of BERT-base, we extract the \texttt{[CLS]} embedding, followed by a dense layer with a \texttt{tanh} activation function, to obtain the final query/passage representation.

\begin{itemize}[topsep=4pt,itemsep=0pt,partopsep=0pt, parsep=4pt]
    \item \textbf{Two-Tower} retrieval models consist of two \textit{independent} BERT-base models: one for each of the query and passage towers, respectively. During inference, we retrieve passages whose embeddings are the most similar to the query embedding.
    \item \textbf{Siamese} retrieval models additionally require that the passage and query BERT models share the same parameters. Siamese networks \cite{bromley1994signature} are popular among tasks that involve quantifying the similarity between comparable items, here a query and a passage.
\end{itemize}

\paragraph{Training Objective} For our retrieval model, we leverage the dot product between query and passage embeddings as a measure of relevance. Following \citet{lee2019latent}, we use sampled Softmax during training.

\section{Datasets}
\label{sec:datasets}
In this section, we describe all the datasets that are used in our experiments. A summary of the datasets is shown in \cref{table:dataset}. 


\paragraph{WikiGQ} (Wikipedia Generated Queries) is the dataset synthesized by our QG model. It consists of 110M synthetically generated queries on 22M passages from our English Wikipedia dataset\footnote{We take the official Wikipedia 2016 database dump and split individual Wikipedia articles into passages with maximum 100 words each, respecting sentence boundaries.}. See \cref{tab:examples} for some examples. This dataset will be open-sourced for research purposes.  

\paragraph{\href{https://ai.google.com/research/NaturalQuestions}{Natural Questions}} \cite{kwiatkowski2019natural} is an end-to-end question answering dataset constructed from English Wikipedia. We convert \textsc{Natural Questions} into a passage retrieval dataset, keeping queries that have long answers. We discard answers that are not regular paragraphs (e.g., tables, lists).  Ultimately, we retain 110,865 of 307,372 queries from the official training set and 4,340 of 7,842 queries from the official development set. Collecting long answers from these queries nets us 84,783 passages. We hold out the 4,340 development queries as the test set.

\paragraph{\href{http://trec-car.cs.unh.edu/}{TREC-CAR}} \cite{dietz2017trec} is a dataset created from English Wikipedia for complex answer retrieval, where the queries are generated by concatenating articles and section titles and the ground truth passages consist of the paragraphs within that section. We use the last segment from the predefined 5 segments of the dataset as our test set. 

\paragraph{\href{https://microsoft.github.io/msmarco/}{MSMARCO}} \cite{nguyen2016ms} comprises over 1M queries sampled from Bing search query logs. Specifically, we use the MSMARCO Passage Ranking (PR) dataset, which contains 8.8M passages from over 3.5M web documents retrieved by Bing. We use the official development set of 55,578 queries as our test set. We also use the official training set of 532,761 queries and their positive passages to train our query generation model. 

\paragraph{\href{https://github.com/shuzi/insuranceQA}{InsuranceQA(v2)}} \cite{feng2015applying} consists of 2,000 evaluation queries on passages sourced from the Insurance Library database.

\paragraph{\href{http://hamedz.ir/resources/}{ANTIQUE}} \cite{hashemi2020antique} is a collection of 2,626 open-domain, non-factoid questions sourced from Yahoo! Answers. We use the provided test set of 200 queries with the corpus of over 400K passages for evaluation. 

\paragraph{\href{http://participants-area.bioasq.org/datasets/}{BioASQ}} \cite{tsatsaronis2012bioasq} Task B consists of 500 English evaluation queries and reference answers constructed by a team of biomedical experts (BioASQ 7b). We collect approximately 20M non-empty article abstracts sourced from the PubMed database as our passage corpus. 

\paragraph{\href{https://github.com/google/retrieval-qa-eval}{ReQA}} \cite{ahmad2019reqa} is a benchmark for evaluating sentence-level answer retrieval models. This benchmark consists of two datasets, \textsc{ReQA SQuAD} and \textsc{ReQA NQ}, created from the official training sets of SQuAD and Natural Questions, respectively.  Each candidate passage is a sentence concatenated with its context passage and a relevant candidate is the answer sentence concatenated with its context passage.

\begin{table*}[t!]
\begin{minipage}{1.0\linewidth}
	\centering
	\footnotesize
\begin{tabu}{@{}llccccccccc@{}}
\toprule
& & \multicolumn{3}{c}{\textsc{Natural Questions}} & \multicolumn{3}{c}{\textsc{TREC-CAR}} & \multicolumn{3}{c}{\textsc{MSMARCO}} \\
\cmidrule(r){3-5} \cmidrule(r){6-8} \cmidrule(r){9-11}
\textbf{Model} & \textbf{Training data} & R@1 & R@10 & R@100 & R@1 & R@10 & R@100 & R@1 & R@10 & R@100  \\
\midrule
BM25  & None & 30.67 & 63.75 & 82.02 & 11.41 & 32.45 & 48.86 & 9.89 & 36.50 & 62.83 \\
\midrule
Two-Tower  &  \textsc{MSMARCO}  & 33.56 & 70.10 & 87.65 & 13.45 & 30.04 & 44.71 & \textit{16.31} & \textit{48.55} & \textit{75.92} \\
Two-Tower  &  \textsc{WikiGQ} & 44.33 & 82.12 & 94.54 & \textbf{18.13} & \textbf{42.46} & 58.49 & 13.38 & 49.49 & 79.93 \\
\midrule
Siamese  & \textsc{MSMARCO}   & 38.52  & 73.17 & 88.36  & 16.85 & 36.41 & 50.88 & \textit{\textbf{18.18}} & \textit{\textbf{53.33}} & \textit{80.46} \\
Siamese &  \textsc{WikiGQ} &\textbf{45.07} & \textbf{82.91} & \textbf{95.32} & 14.91 & 40.20 & \textbf{58.87} & 12.98 & 50.74 & \textbf{81.38} \\
\midrule\midrule
Siamese &  \textsc{Official data} & \textit{40.78} & \textit{80.77} & \textit{94.96} & \textit{22.10} & \textit{48.18} & \textit{67.47} & \textit{18.18} & \textit{53.55} & \textit{80.46} \\
\midrule
\toprule
&  & \multicolumn{3}{c}{\textsc{InsuranceQA}} & \multicolumn{3}{c}{\textsc{ANTIQUE}} & \multicolumn{3}{c}{\textsc{BioASQ}} \\
\cmidrule(r){3-5} \cmidrule(r){6-8} \cmidrule(r){9-11}
\textbf{Model} & \textbf{Training data} & R@1 & R@10 & R@100 & R@1 & R@10 & R@100 & R@1 & R@10 & R@100  \\
\midrule
BM25      & None             & 22.41 & 50.61 & 78.98 & 3.25 & 17.59 & 42.16 & \textbf{20.28} & \textbf{45.32} & \textbf{71.83} \\
\midrule
Two-Tower & \textsc{MSMARCO}       & 19.74 & 48.21 & 75.56 & 4.79 & 16.23 & 36.20 & 3.25 & 10.53 & 22.34 \\
Two-Tower & \textsc{WikiGQ}        & 20.27 & 50.29 & 78.71 & 5.15 & 22.65 & 46.51 & 9.22 & 24.24 & 43.50 \\
Two-Tower & \textsc{WikiGQ+FT(GQ)} & 30.20 & 67.27 & 91.37 & 5.90 & 22.45 & 45.83 & 14.39 & 33.46 & 55.45 \\
\midrule
Siamese   & \textsc{MSMARCO}       & 23.19 & 52.52 & 79.72 & 5.05 & 19.04 & 40.75 & 6.41 & 14.64 & 31.79 \\
Siamese   & \textsc{WikiGQ}        & 23.24 & 54.65 & 81.76 & 4.15 & 17.35 & 43.06 & 12.30 & 30.60 & 52.17 \\
Siamese   & \textsc{WikiGQ+FT(GQ)} & \textbf{31.47} & \textbf{68.85} & \textbf{92.05} & \textbf{6.09} & \textbf{23.38} & \textbf{49.01} & 15.91 & 37.30 & 61.77 \\
\midrule\midrule
Siamese &  \textsc{Official data} & \textit{30.82} & \textit{67.73} & \textit{92.88} & \textit{3.89} & \textit{18.54} & \textit{41.83} & \textit{6.50} & \textit{19.77} & \textit{36.90} \\
\bottomrule
\end{tabu}
\end{minipage}
\caption{Detailed results on zero-shot performance. Rows with `\textsc{MSMARCO}' as `training data' use the official \textsc{MSMARCO} PR training set to train the models. Rows with `\textsc{WikiGQ}' use our synthetic \textsc{WikiGQ} dataset as training data. Rows with `\textsc{WikiGQ+FT(GQ)}' denotes further finetuning on domain-specific synthetic data after pretraining on \textsc{WikiGQ}. Models trained on official training sets, including rows corresponding to \textsc{Official data}, are in \textit{italics}. Best zero-shot numbers are in \textbf{bold}.
Numbers are in percent (\%).}
\label{table:zeroshot}
\end{table*}

\begin{table}[h!]
\begin{minipage}{1.0\linewidth}
	\centering
	\footnotesize

\begin{tabu}{@{}llccc@{}}
\toprule
 &  & \multicolumn{3}{c}{\textbf{Average Recall}} \\
\cmidrule(r){3-5} 
\textbf{Model} & \textbf{Training data} & R@1 & R@10 & R@100 \\
\midrule
BM25      & None             & 16.32 & 41.04 & 64.45 \\
\midrule
Two-Tower & \textsc{MSMARCO}       & 15.18 & 37.28 & 57.06 \\
Two-Tower & \textsc{WikiGQ}        & 18.41 & 45.21 & 66.95 \\
\midrule
Siamese   & \textsc{MSMARCO}       & 18.03 & 41.52 & 61.99 \\
Siamese   & \textsc{WikiGQ}        & \textbf{18.77} & \textbf{46.08} & \textbf{68.76} \\
\bottomrule
\end{tabu}
\end{minipage}
\caption{Average results on zero-shot performance. Numbers are in percent(\%).}
\label{table:avg}
\end{table}

\section{Implementation Details}
\label{sec:implementation}

\paragraph{Generation} We begin by finetuning the pretrained BART-large (374M parameters) on MSMARCO, generating queries given a passage. We train BART for 5 epochs with a batch size of 96 and a learning rate of 3e-5 (warmup ratio $0.1$). Once the BART model is trained, we perform generation over English Wikipedia passages to generate 10 synthetic queries for each passage, using nucleus sampling \cite{holtzman2019curious} with $p=0.95$, and retaining the top-5 queries based on likelihood scores. The resulting dataset consists of 110M synthetic query-passage pairs, which we call \textsc{WikiGQ}. 

\paragraph{Pretraining} We initialize our Siamese/two-tower retrieval models with BERT-base (110M parameters). We then pretrain our retrieval models on \textsc{WikiGQ}. The pretraining is done on 32 Nvidia V100 GPUs with a batch size of 100 per GPU for 10 epochs. We set a max learning rate of 5e-5 with warmup ratio of $0.1$. The pretraining takes about 4 days for the Siamese model and 6 days for the Two-Tower model. 

For our baseline models trained on the MSMARCO official training set, we train for 10 epochs on 8 GPUs with a batch size of 50 per GPU and max learning rate of 3e-5 with warmup ratio of $0.1$. Half-precision training\footnote{\url{https://github.com/nvidia/apex}} is enabled in all of our experiments. 

For our Inverse Cloze Task (ICT) baselines, we follow the process defined in \citet{lee2019latent}. For each passage, we randomly sample one sentence as a query and remove it from the passage $90\%$ of the time. To ensure fair comparison, we train on the ICT data, where data is generated on-the-fly, for 5 times as many epochs as on the QG synthetic data in order to match the number of training iterations. Specifically, we train our models on QG synthetic data for 10 epochs (50 epochs for ICT) on 8 GPUs with a batch size of 50 per GPU and learning rate 3e-5. 

\paragraph{Finetuning} We use the same BART QG model (trained on \textsc{MSMARCO}) to generate domain-specific synthetic data (we do this for several datasets that are not derived from English Wikipedia: \textsc{InsuranceQA, ANTIQUE, BioASQ}). When finetuning our retrieval models on domain-specific synthetic data, we set max learning rate to 3e-5 with warmup ratio of $0.1$ and finetune on 8 GPUs with a batch size of 50 per GPU for 10 epochs (5 epochs for \textsc{BioASQ}).

\section{Experimental Results}
\label{sec:results}
In this section, we present our main results on zero-shot retrieval. Specifically, we consider the following models and baselines:
\begin{itemize}[topsep=2pt,itemsep=0pt,partopsep=0pt, parsep=2pt]

    \item \textbf{BM25}: a classical IR method based on lexical term-matching. We consider BM25 as a strong baseline for an unsupervised retrieval system. We leverage the implementation from Elasticsearch\footnote{\url{https://www.elastic.co/}} with the default settings.
    
    \item \textbf{Siamese/Two-Tower + MSMARCO}: Because we use MSMARCO to train our BART query generator, we consider a Siamese/Two-Tower model trained on MSMARCO as a reasonable baseline to compare against.

    \item \textbf{Siamese/Two-Tower + WikiGQ}: We train both the Siamese and Two-Tower models on the \textsc{WikiGQ} dataset. 
    
    \item \textbf{Siamese/Two-Tower + WikiGQ + domain-specific synthetic data}: For datasets that are not based on English Wikipedia, we finetune on the synthetic data additionally generated from said datasets.
    
\end{itemize}

The results are shown in Tables \ref{table:zeroshot} \& \ref{table:avg}. We find that, in most cases, our models trained on the synthetic \textsc{WikiGQ} data outperforms BM25 and the models trained on \textsc{MSMARCO}. 
In particular, for datasets such as \textsc{Natural Questions} and \textsc{TREC-CAR}, which are based on English Wikipedia, the recall from the models trained on \textsc{WikiGQ} is about $20\%$ - $50\%$ relatively higher than the BM25 baselines. For non-Wikipedia datasets such as \textsc{InsuranceQA} and \textsc{ANTIQUE}, we obtain a significant boost in performance through unsupervised domain adaptation. 

One notable exception is the \textsc{BioASQ} dataset on which our neural retrieval models fail to outperform the BM25 baseline. One possible reason lies in the construction of the \textsc{BioASQ} dataset itself.\footnote{Please see details in \url{http://participants-area.bioasq.org/general_information/Task7b/}} In this dataset, human annotations are built from candidates retrieved via term-matching with optional boosting tags \cite{tsatsaronis2012bioasq}. Furthermore, the annotation depth is relatively shallow (approximately 10-60 articles per query) whereas the total number of articles is around 20M. We believe that this annotation process favors lexical term-matching systems like BM25. 

Comparing the results from the Siamese and Two-Tower models, we find that the Siamese model generally outperforms the Two-Tower model, indicating that sharing parameters across encoders is helpful. This is consistent with observations from \citet{das2016together}.

Finally, for each dataset, we report the performance of the Siamese model trained with official training sets. The results are reported in \textit{italics} in the bottom rows of each section in \cref{table:zeroshot}. Remarkably, we find that in 4 out of 6 datasets, Siamese models trained purely on synthetic data can already outperform the models trained on the official training sets. In particular, for \textsc{ANTIQUE} and \textsc{BioASQ}, which have relatively small training sets, Siamese models trained on synthetic data improve Recall@1 by over 50\% and 140\%, respectively. This further demonstrates that QG can be helpful when only a limited amount of labeled data is available. 
\section{Ablations}
\label{sec:ablations}
\begin{figure}[h!]
    \centering
    \includegraphics[width=1.0\columnwidth]{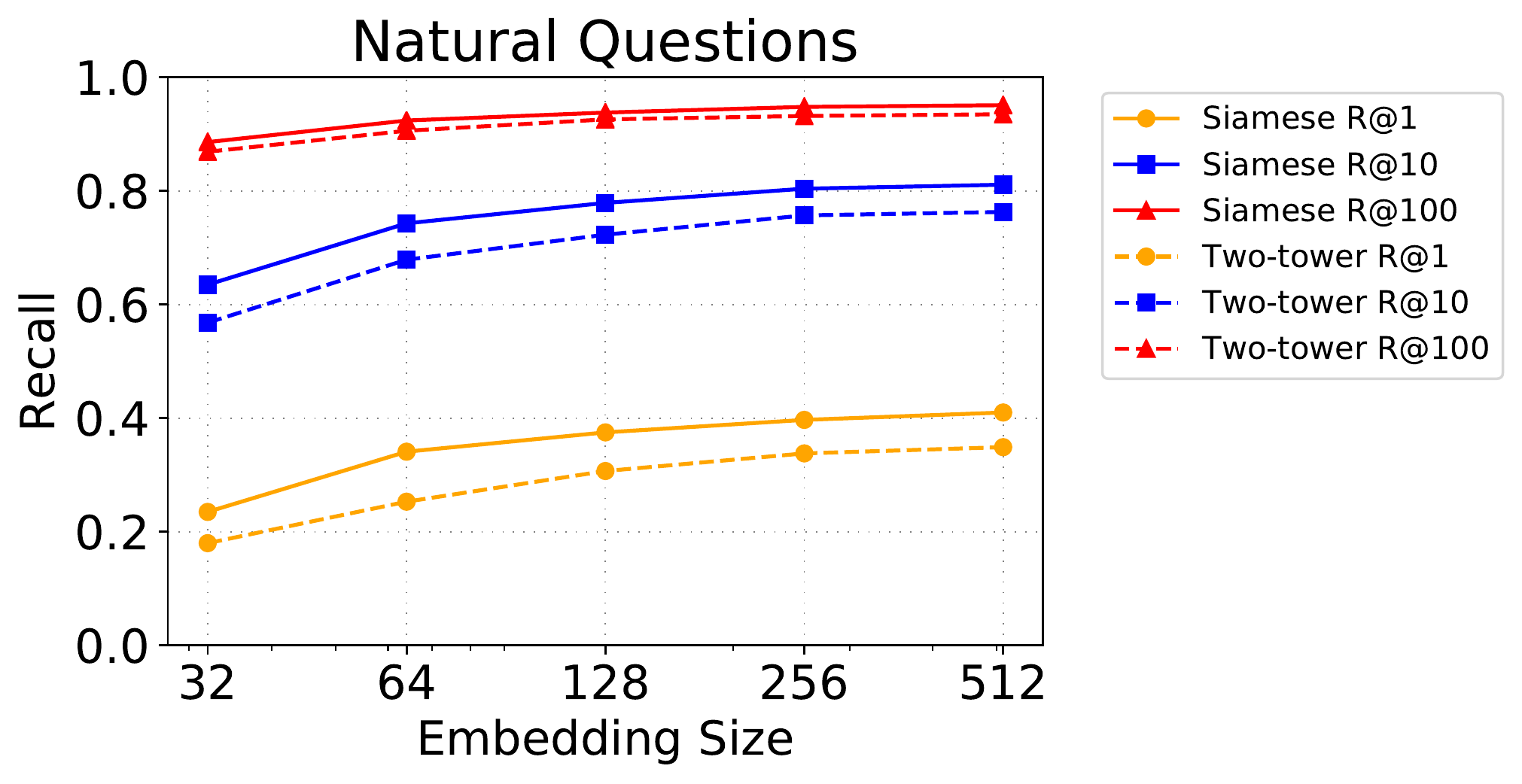}
    \caption{Comparison between Siamese and Two-Tower networks with different embedding sizes.}
    \label{fig:ablation_embedding}
\end{figure}

\paragraph{Embedding Size} We consider two variations of retrieval model: the Two-Tower model (without shared weights) and the Siamese model (with shared weights). 
For each variation we test various embedding sizes for the BERT-base encoders. For simplicity, we perform our ablations on the (non-synthetic) \textsc{Natural Questions} dataset. \cref{fig:ablation_embedding} shows a monotonic increase in performance as the embedding size grows larger, slowly plateauing around an embedding size of 512. 

\begin{table}[h!]
\begin{minipage}{1.0\linewidth}
	\centering
	\footnotesize
\begin{tabu}{@{}llccc@{}}
\toprule
\multirow{2}{*}{\shortstack{\textbf{Generation}\\ \textbf{Method}}}& \multirow{2}{*}{\textbf{\# pairs}} & \multicolumn{3}{c}{\textsc{Natural Questions}}  \\
& & R@1 & R@10 & R@100 \\
\midrule
\underline{\textbf{Siamese}} & & & & \\
ICT  & N/A & 18.85 & 48.26 & 73.54 \\
Beam    &      422,309    & 40.15  & 77.40 & 91.51 \\
Nucleus ($p=0.95$) &  363,303 &\textbf{40.97} & \textbf{79.36} & \textbf{93.16} \\
\midrule
\underline{\textbf{Two-tower}} & & & & \\
ICT      & N/A & 13.72 & 40.13 & 67.36 \\
Beam      &  422,309  & 35.97 & 73.19 & 89.09 \\
Nucleus ($p=0.95$)&  363,303 & \textbf{37.58} & \textbf{75.56} & \textbf{91.30} \\
\bottomrule
\end{tabu}
\end{minipage}
\caption{Comparison between ICT and synthetic pretraining. `Beam' refers to beam search with beam width 5 as the generator decoding strategy. `Nucleus' refers to taking the top-5 samples from the 10 independent sequences generated via nucleus sampling with $p=0.95$. Numbers are in percent (\%).  }
\label{table:ablation_decoding}
\end{table}
\begin{table}[t!]
\begin{minipage}{0.99\linewidth}
	\centering
	\footnotesize
	
    \begin{tabu}{p{\textwidth}}
    \toprule
    \textbf{Original Query-Passage Pair}\\
    \midrule
    \textbf{Passage}: The BMW E70 is the second generation X5 Sports Activity Vehicle ( SAV ) . It replaced the BMW X5 ( E53 ) in November 2006 . The second generation X5 features many new technological advancements including BMW 's iDrive system as standard equipment and , for the first time in a BMW , an optional third row seat raising passenger capacity to seven .\\
    \\
    \textbf{Original Query}: when did the new shape bmw x5 come out \\
    \midrule
    \textbf{Synthetic Queries (Beam Search)}: \\
    \midrule
    what is bmw x5 \\
    what is bmw e70 \\
    what year did bmw x5 come out \\
    what year did the bmw x5 come out \\
    what year was the bmw x5 made \\
    \midrule
    \textbf{Synthetic Queries (Nucleus Sampling)}: \\
    \midrule
    what year did bmw x5 come out  \\
    what year was the bmw x5 sports activity vehicle \\
    what is bmw sav \\
    how many people does a bmw x5 seat \\
    what year was bmw x5 sports activity vehicle (sav) made \\
    \bottomrule
    \end{tabu}
\end{minipage}
\caption{Examples of synthetic queries on \textsc{Natural Questions} passages. Synthetic queries from nucleus sampling not only are similar to real queries, but also cover significantly more passage content.}
\label{tab:synthetic}
\end{table} 
\begin{table*}[h]
\begin{minipage}{1.0\linewidth}
	\centering
	\footnotesize
\begin{tabu}{@{}llcccccccccc@{}}
\toprule
& & \multicolumn{5}{c}{\textsc{ReQA SQuAD}} & \multicolumn{5}{c}{\textsc{ReQA NQ}}  \\
\cmidrule(r){3-7} \cmidrule(r){8-12} 
\textbf{Model}    & \textbf{Pretraining} & {R@1}   & {R@5}   & {R@10}  & {R@50}  & {R@100} & {R@1}   & {R@5}   & {R@10}  & {R@50}  & {R@100} \\
\midrule
USE-QA    & -           & 43.9  & 65.6  & 72.7  &  -    & -    & 14.7  & 31.7  & 39.1  & -     & -     \\
Two-Tower$^\dagger$ & \textsc{ICT+BFS+WLP} & 37.43 & 61.48 & 70.18 & 85.37 & 89.85 & 17.31 & 43.62 & 55.00 & 76.59 & 82.84 \\
\midrule
BM25  & -           & \textbf{58.51} & \textbf{76.80} & \textbf{82.14} & 90.45 & 92.72 & 18.06 & 42.23 & 52.26 & 70.71 & 76.52 \\
Two-Tower & \textsc{WikiGQ}      & 41.29 & 67.59 & 76.03 & 89.10 & 92.51 & 21.71 & 55.55 & 68.35 & 86.65 & 90.83 \\
Siamese   & \textsc{WikiGQ}      & 46.53 & 72.52 & 80.27 & \textbf{91.42} & \textbf{94.19} & \textbf{21.88} & \textbf{56.48} & \textbf{69.67} & \textbf{87.93} & \textbf{91.88} \\
\bottomrule
\end{tabu}
\end{minipage}
\caption{Zero-short performance on ReQA datasets. USE-QA is reported from \cite{ahmad2019reqa} and Two-Tower$^\dagger$ is reported from \cite{chang2020pre}. Since \citet{chang2020pre} did not report zero-shot performance, we report their numbers with $1\%/99\%$ training/test split setting. Numbers are in percent (\%).}
\label{table:reqa}
\end{table*}

\paragraph{Pretraining Methodology} Recently, several pretraining tasks have been proposed to improve the performance of embedding-based neural retrieval. ICT was first introduced by \citet{lee2019latent} with strong performance on open-domain QA tasks. Subsequently, \citet{chang2020pre} introduced two new pretraining tasks, Body First Selection (BFS) and Wiki Link Prediction (WLP), in addition to ICT. In this ablation study, we directly compare these pretraining methods against our own. 

Beginning with the \textsc{Natural Questions} dataset, we compare models trained with ICT against our own model, trained with synthetic queries. When synthesizing queries, we consider two decoding techniques --- beam search and nucleus sampling \cite{holtzman2019curious}. With beam search decoding, using a beam width of 5, we generate 5 queries for each passage. Using nucleus sampling, with $p=0.95$, we generate 10 independent samples, selecting the top 5 queries based on the likelihood scores. After removing duplicate queries, the total number of generated query-passage pairs  
is shown in \cref{table:ablation_decoding}.

We find that both Siamese and Two-Tower models, pretrained with synthetic queries, significantly outperform their ICT counterparts.
We hypothesize that models trained on ICT (where the model attempts to retrieve a passage, given a sentence from that passage) may suffer from this inductive bias when used to perform ranking (where the model must retrieve a passage, given a user query). 
On the other hand, synthetic queries from our generator are very similar to real user queries. To illustrate this point further, we provide examples in \cref{tab:synthetic}. 

Additionally, we find that training with synthetic queries generated via nucleus sampling outperforms training with queries generated with beam search. While beam search decoding generates strictly more training pairs than nucleus sampling (nucleus sampling can generate duplicates across $k$ independent runs), nucleus sampling tends to result in more diverse queries \cite{holtzman2019curious}. We show examples in \cref{tab:synthetic}.

For completeness, we additionally compare our model against the ICT, BFS and WLP pretraining tasks in conjunction. 
Following \citet{chang2020pre}, we use all of English Wikipedia during pretraining for both synthetic query generation (i.e., \textsc{WikiGQ} as described in \cref{sec:implementation}) and evaluate the resulting models on the ReQA benchmark.

We report our zero-shot retrieval results on the BM25 baseline and models trained on \textsc{WikiGQ}. A summary of the results are shown in \cref{table:reqa}\footnote{Our BM25 baseline significantly outperforms those reported in \citet{ahmad2019reqa,chang2020pre}. As described earlier, our BM25 results are obtained from Elasticsearch using default settings. We have communicated with the authors from \citet{ahmad2019reqa} and they were able to reproduce our results.}. As we can see, on both datasets, models pretrained on our synthetic \textsc{WikiGQ} corpus significantly outperform their counterparts pretrained with ICT+BFS+WLP. Notably, the Two-Tower model trained on ICT+BFS+WLP is actually finetuned with $1\%$ of the original training data after \textsc{ICT+BFS+WLP} pretraining. However, our models are trained only on the synthetic \textsc{WikiGQ} data (no real data was used). These results indicate that pretraining with \textsc{WikiGQ} is more favorable than ICT, BFS and WLP for retrieval. Finally, our models pretrained on synthetic data outperform BM25 on \textsc{ReQA NQ} by a large margin. However, our model is unable to consistently outperform BM25 on \textsc{ReQA SQuAD}. We believe that this is due to certain characteristics of the SQuAD dataset. Similar observations have been made in \citet{lee2019latent,karpukhin2020dense}\footnote{They present two possible reasons. First, the high lexical overlap between SQuAD questions and passages due to the fact that the questions were created by the annotators after seeing the passages. Second, data was collected from only 500+ Wikipedia articles which makes the data highly correlated and suboptimal for models trained on i.i.d samples.}.

\begin{table}[hb]
\begin{minipage}{1.0\linewidth}
	\centering
	\footnotesize

\begin{tabu}{@{}lccc@{}}
\toprule
\multirow{2}{*}{\textbf{Model}} & \multicolumn{3}{c}{\textbf{Recall}}  \\
&   R@1 & R@10 & R@100 \\
\midrule
\underline{\textbf{Siamese}} & \multicolumn{3}{c}{\textsc{\textsc{Natural Questions}}} \\
\ \ FT w/o pretraining & 40.78 & 80.77 & 94.96\\
\ \ FT w/ \textsc{WikiGQ} pretraining & \textbf{48.57} & \textbf{88.29} & \textbf{97.30} \\
\midrule\midrule
\underline{\textbf{Siamese}} & \multicolumn{3}{c}{\textsc{\textsc{InsuranceQA}}} \\
\ \ FT w/o pretraining  & 30.82 & 67.72 & 92.88  \\
\ \ FT w/ \textsc{WikiGQ} pretraining &  \textbf{34.33} & \textbf{73.46} & \textbf{95.71} \\
\bottomrule
\end{tabu}

\end{minipage}

\caption{Abalation study on Siamese models, with and without pretraining on \textsc{WikiGQ}. Here we use the official training set of \textsc{Natural Questions} and \textsc{InsuranceQA} for finetuning. Numbers are in percent (\%).}
\label{table:ablation_pretraining}
\end{table}

\begin{figure*}[ht!]
    \centering
    \includegraphics[width=1.8\columnwidth]{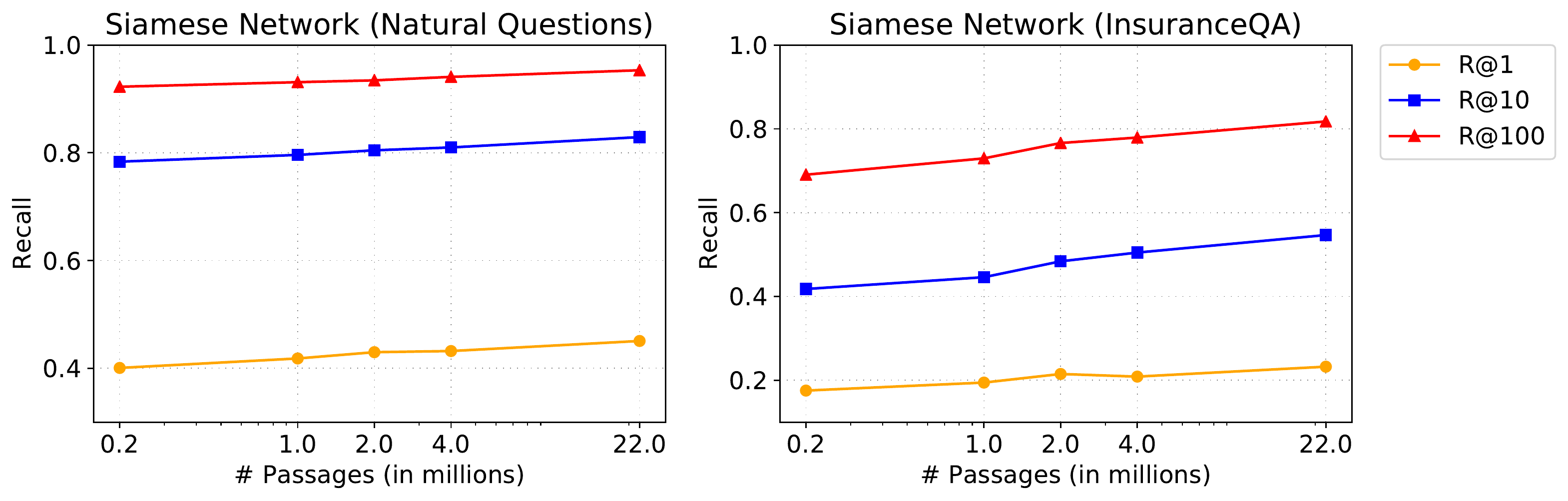}
    \caption{Ablations on amount of passages used for pretraining.}
    \label{fig:ablation_pretraining_amount}
\end{figure*}
\paragraph{With or Without Pretraining} We study the impact of \textsc{WikiGQ} pretraining when we can finetune models on hand-labeled training data. Specifically, we finetune the Siamese models with and without pretrained on \textsc{WikiGQ} on the official training data of two datasets, 
\textsc{Natural Questions} and \textsc{InsuranceQA}. 
Our results show that pretraining with \textsc{WikiGQ} improves performance for both Wikipedia-based dataset (\textsc{Natural Questions}) and out-of-domain dataset (\textsc{InsuranceQA}). We detail our results in \cref{table:ablation_pretraining}.

\paragraph{Data Efficiency with Pretraining}
Finally, we study the sample efficiency of synthetic pretraining with \textsc{WikiGQ}. We evaluate on two datasets, \textsc{Natural Questions} (Wikipedia-based) and \textsc{InsuranceQA} (non-Wikipedia-based). Specifically, we train the Siamese model on synthetic data generated with various fractions of English Wikipedia. Our results show that performance improves monotonically with synthetic dataset size on both \textsc{Natural Questions} and \textsc{InsuranceQA}. However, we witness diminishing returns, and training on 22M passages only confers a slight improvement over 4M passages. The results are shown in \cref{fig:ablation_pretraining_amount}.

\section{Related Work}
Recent work has demonstrated the effectiveness of embedding-based neural retrieval models on large document corpora. \citet{lee2019latent} first introduced the Inverse Cloze Task (ICT) as a pretraining task for neural retrieval models, demonstrating improved performance over BM25 for open-domain question answering (QA) tasks. \citet{chang2020pre} proposed additional pretraining tasks and showed improved performance in the ReQA benchmark \cite{ahmad2019reqa}. However, none of these works have demonstrated effectiveness in the zero-shot setting.

Along this vein, leveraging generative models to produce synthetic data has been previously explored. \citet{du2017learning} first applied the seq2seq model for automatic question generation from text in reading comprehension. \citet{tang2017question,sachan2018self} proposed to jointly train QG and QA models to improve QA performance.
\citet{lewis-etal-2019-unsupervised} trained an unsupervised sequence-to-sequence model to generate natural questions from cloze content.
\citet{alberti2019synthetic} generated queries and answers by finetuning BERT on extractive subsets of SQuAD 2.0 and Natural Questions and employing a sequence-to-sequence model for query generation. \citet{puri2020training} first demonstrated that a QA model trained on purely synthetic questions and answers can outperform models trained on human-labeled data on SQuAD1.1. None of these have studied using QG for passage retrieval tasks.

A contemporaneous work uses synthetic queries for domain adaptive neural retrieval \cite{ma2020zero}. Our work additionally considers large-scale pretraining with synthetic queries, additional evaluation datasets, and careful ablations.

\section{Conclusions}

In the paper, we proposed synthetic query generation for improving the performance of embedding-based neural retrieval models in the zero-shot setting. We synthesize \textsc{WikiGQ}, a large-scale synthetic retrieval dataset from English Wikipedia. Leveraging \textsc{WikiGQ}, our retrieval models are able to outperform other pretraining strategies such as ICT while also exhibiting superior zero-shot performance on multiple datasets from various domains. Finetuning on the domain-specific synthetic data further improves performance. 

Future work on synthetic query generation should address questions about data quality. For example, methods for intelligent filtering to reduce the number of unanswerable queries may lessen the noise present in synthetic data. Decreasing the amount of redundant data through improvements to decoding strategies may further improve training efficiency. Finally, practitioners should consider leveraging better models and larger datasets for training both generation and retrieval models.

\bibliographystyle{acl_natbib}
\bibliography{anthology,emnlp2020}

\newpage

\end{document}